\pdfoutput=1
\documentclass[
amsfonts,amssymb,amsmath,
nofootinbib,superscriptaddress, twocolumn
]{revtex4-1}

\usepackage{sastyle} 
\usepackage{amsmath,bm,mathtools,physics}
\usepackage[utf8]{inputenc}

\numberwithin{equation}{section}

\usepackage{tikz}
\usepackage{graphicx}
\usepackage{comment}
\graphicspath{ {/} }

\newcommand{\arr}[1]{%
	\parbox{#1}{\tikz{\draw[->](0,0)--(#1,0);}}
}

\usepackage{color}

\usepackage{color}
\definecolor{darkblue}{rgb}{0.1,0.1,.7}
\usepackage[colorlinks, linkcolor=darkblue, citecolor=darkblue, urlcolor=darkblue, linktocpage]{hyperref} 
\begin{document}
\title{On duality of color and kinematics in (A)dS momentum space}
\author{Soner Albayrak}
\affiliation{Department of Physics, Yale University, New Haven, CT 06511}
\email{soner.albayrak@yale.edu}
\author{Savan Kharel} 
\affiliation{Department of Physics, University of Chicago, Chicago, IL 60637}
\email{skharel@uchicago.edu}
\author{David Meltzer} 
\affiliation{Walter Burke Institute for Theoretical Physics, California Institute of Technology, Pasadena, CA 91125}
\email{dmeltzer@caltech.edu}

\begin{abstract}

We explore color-kinematic duality for tree-level AdS/CFT correlators in momentum space. 
We start by studying the bi-adjoint scalar in AdS at tree-level as an illustrative example. 
We follow this by investigating two forms of color-kinematic duality in Yang-Mills theory, the first for the integrated correlator in AdS$_4$ and the second for the integrand in general AdS$_{d+1}$.
For the integrated correlator, we find color-kinematics does not yield additional relations among $n$-point, color-ordered correlators.
To study color-kinematics for the AdS$_{d+1}$ Yang-Mills integrand, we use a spectral representation of the bulk-to-bulk propagator so that AdS diagrams are similar in structure to their flat space counterparts.
Finally, we study color KLT relations for the integrated correlator and double-copy relations for the AdS integrand.
We find that double-copy in AdS naturally relates the bi-adjoint theory in AdS$_{d+3}$ to Yang-Mills in AdS$_{d+1}$.
We also find a double-copy relation at three-points between Yang-Mills in AdS$_{d+1}$ and gravity in AdS$_{d-1}$ and comment on the higher-point generalization.
By analytic continuation, these results on AdS/CFT correlators can be translated into statements about the wave function of the universe in de Sitter.

	\end{abstract}
\date{\today}
\maketitle

\section{Introduction}
The study of scattering amplitudes and on-shell observables in quantum field theory has revealed new mathematical structures and symmetries which are obscured in off-shell, Lagrangian formulations \cite{Elvang:2013cua}. 
The duality between color and kinematics, and the associated double-copy relations, are prominent examples that give fundamentally new insights into the perturbative structure of quantum field theory \cite{Bern:2010ue}. 
These ideas indicate that the dynamics of gauge theories and gravity, when they are both weakly coupled, are governed by the same kinematical building blocks.
Additionally, these ideas have lead to novel computations for loop-level graviton amplitudes, gravitational wave patterns, and string theory amplitudes. Color-kinematic duality and double-copy have also been applied to theories with seemingly no relation to gauge or gravity theories. We refer the reader to \cite{Bern:2019prr} for a recent, comprehensive review of these topics. 

Concurrent to these advances in flat space scattering amplitudes, there has also been an intense focus on the study of holographic correlators. The most concrete example of holography has been formulated in asymptotically anti-de Sitter (AdS) spacetimes \cite{Maldacena:1997re}. While there is no notion of an S-matrix in AdS, one can consider AdS scattering experiments that are dual to the correlation functions of a conformal field theory (CFT) living on its boundary. In this context, the terms ``AdS amplitude'' and ``CFT correlator'' are typically used interchangeably.
There are now a variety of methods to compute holographic correlators, including ideas from Mellin space \cite{Penedones:2010ue,Fitzpatrick:2011ia,Rastelli:2016nze}, the conformal bootstrap \cite{Heemskerk:2009pn,Alday:2017gde}, and harmonic analysis in AdS \cite{Costa:2014kfa,Sleight:2017fpc}.  In this work, we will build on recent developments concerning CFT correlation functions in momentum space \cite{Bzowski:2013sza, Bzowski:2015pba, Bzowski:2018fql, Bzowski:2015pba, Bzowski:2019kwd, Bzowski:2020kfw, Isono:2018rrb, Isono:2019wex, Coriano:2013jba, Coriano:2018bbe, Anand:2019lkt, Gillioz:2019lgs, Farrow:2018yni, Nagaraj:2019zmk, Nagaraj:2020sji, Jain:2020rmw, Raju:2010by, Raju:2011mp,Raju:2012zr, Raju:2012zs, Meltzer:2020qbr, Albayrak:2018tam, Albayrak:2019asr, Albayrak:2019yve, Albayrak:2020isk, Albayrak:2020bso} in order to understand how ideas from flat space amplitudes can be imported to curved spacetimes. 

One motivation to study holographic correlators in momentum space comes from their close connection to the wave function of the universe \cite{Maldacena:2002vr, Harlow:2011ke,Ghosh:2014kba}, which can be used to compute late-time cosmological correlators. Inspired by the modern amplitudes program, there is an ongoing systematic program to compute de Sitter invariant correlators, which is known as the \emph{cosmological bootstrap} \cite{Arkani-Hamed:2018kmz,Baumann:2019oyu,Baumann:2020dch,Sleight:2019hfp,Sleight:2020obc, Pajer:2020wxk, Pajer:2020wnj}. Interestingly, holographic and cosmological correlators possess a total energy singularity when the norms of all the momenta sum to zero. The coefficient of this singularity is exactly the scattering amplitude for the same process in flat space \cite{Maldacena:2011nz,Raju:2012zr}. In other words, holographic and cosmological correlators contain within them information about flat space amplitudes.

It is then natural to wonder if one can generalize the rich structure of color-kinematic duality and double-copy to AdS and cosmological correlators. Color-kinematic duality implies that flat space scattering amplitudes can be arranged in such a way that the kinematic numerators of the scattering amplitude have the same algebraic properties as the color factors. 
That is, whenever the color factors of an amplitude obey a Jacobi identity, the corresponding kinematic factors can also be chosen such that they obey the same relation. 
The double-copy construction, which relates Yang-Mills amplitudes to gravity, then corresponds to replacing the color factors of a Yang-Mills amplitude with the corresponding, color-kinematic obeying numerators. 
Since color-kinematics has lead to computational and conceptual advances in flat space scattering amplitudes, it is natural to hope that similar advances can be made for curved space correlators. 

In this work, we take the initial steps in generalizing and testing color-kinematic duality for AdS/CFT correlators, or equivalently for the cosmological wave function. 
We will propose two different formulations of color-kinematics in AdS momentum space. 
The first method corresponds to imposing this duality on the full, integrated correlator. In the examples considered here, we find it is always possible to choose a set of numerators such that color-kinematics holds for the integrated correlator. 
The second method corresponds to imposing color-kinematics directly on the AdS integrand. This is inspired by recent work on scattering in a plane-wave background \cite{Adamo:2018mpq} and on the study of celestial amplitudes \cite{Casali:2020vuy}.
We also comment on its connection to double-copy.

This note is organized as follows. 
In section \ref{sec:bi-adjoint} we will study the scalar bi-adjoint theory in AdS at tree-level. 
This will serve as a simple example to illustrate how color-kinematics works in AdS and how the BCJ relations are modified. 
In section \ref{sec:YM}, we study color-kinematics for Yang-Mills four-point functions in AdS, both at the level of the integrated correlator and the integrand. 
Finally, in section \ref{sec:KLT} we comment on the color KLT relations, which connect the bi-adjoint and Yang-Mills theories, and the double-copy relations between Yang-Mills and gravity at three and four-points in AdS.

\textit{Note: After this work was completed,
\cite{Armstrong:2020woi} appeared which partially overlaps with our results.}
\section{Bi-Adjoint Scalar}
\label{sec:bi-adjoint}
To start, let us recall that the bi-adjoint scalar theory consists of scalars $\f^{aA}$ which are charged under two different $SU(N)$ global symmetries (see \cite{Du:2011js, Cachazo:2013iea, White:2016jzc} and references therein). We will use lowercase and capital Latin letters to distinguish the two groups. The action for this theory is simple and takes the following form:  
\be 
	S_{\text{bi-adjoint}}=-\int d^{d+1}x\sqrt{g}\bigg(\frac{1}{2}(\nabla_{\mu}\f^{aA})(\nabla^{\mu}\f_{aA})\\+\frac{\xi R}{2}\f_{aA}\f^{aA}-\lambda f_{abc}f_{ABC}\f^{aA}\f^{bB}\f^{cC}\bigg)~,
\ee 
where $R$ is the Ricci scalar.
We have added the Ricci scalar with an arbitrary $\xi$ to be general and will later fix it so the scalar is conformally coupled.
In AdS\footnote{We will focus on AdS computations for concreteness in the rest of the note, although our results can equally be interpreted in terms of the dS wave function with suitable analytic continuation \cite{Maldacena:2002vr}.}, the metric $g$ for the Poincar\'e patch is
\be 
	ds^2=\frac{dz^{2}+\eta_{\mu\nu}dx^{\mu}dx^{\nu}}{z^2}~.
\ee 
Here we take $\eta^{\mu\nu}$ to be the mostly plus, flat space metric.
The scalar $\f^{aA}$ is dual to a boundary operator\footnote{The bulk theory does not include gravity, so the boundary theory does not have a stress-tensor and will be non-local.} $\mathcal{O}^{aA}$ with conformal dimension $\Delta=d/2+\nu$ where
\be 
	\nu&=\frac{1}{2} \sqrt{d (d-4 (d+1) \xi )}~.
\ee 
A particularly simple case to study is the conformally coupled scalar, which corresponds to,
\be 
	\xi_{c}=\frac{d-1}{4d}~,
\ee 
or $\nu=1/2$.

\begin{figure}
	\centering
	$\begin{aligned}
		\includegraphics[scale=.6]{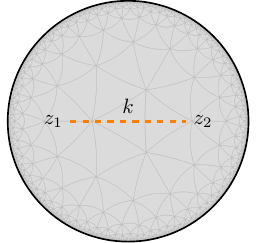}
	\end{aligned}=G^{\text{sc}}(k,z_1,z_2)$
	$\begin{aligned}
		\includegraphics[scale=0.6]{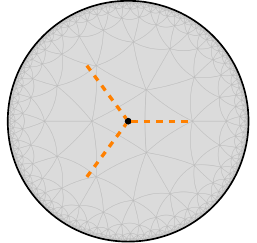}
	\end{aligned}=i\lambda$
	\caption{\label{fig: Feynman rules for bi-adjoint scalar} AdS Feynman rules for $\f^3$ theory.}
\end{figure}

Next, we need the bulk-to-bulk propagator for scalars in AdS:
\be 
	G^{\text{sc}}_{\nu}(k,z_1,z_2)=-i(z_1z_2)^{\frac{d}{2}}\int\limits_{0}^{\infty}\frac{dp^2}{2}\frac{\cJ_{\nu}(pz_1)\cJ_\nu(pz_2)}{\bk^{2}+p^{2}-i\epsilon}~,
\ee 
where $\cJ$ is the modified Bessel function of the first kind and $k\equiv\abs{\bk}= \sqrt{\bk^2}$ is the norm of the boundary momenta $\bk$. When computing correlators, we will always take the external momenta to be spacelike. When we set $\nu=1/2$, the $p$ integral can be computed in closed form, but we find the $p$ integral representation of the propagator simplest for practical computations. By taking one point to the boundary, we find the scalar bulk-to-boundary propagator:
\be 
	K^{\text{sc}}_{\nu}(k,z)= -i\frac{1}{2^{\nu}\Gamma(1+\nu)}z^{\frac{d}{2}}k^{\nu}\cK_{\nu}(kz)~,
\ee
where $\cK$ is the modified Bessel function of the second kind. Finally, to each interaction vertex we have the factor $i\lambda f_{abc}f_{ABC}$. 

\subsection{Test case: AdS$_6$}
\label{sec:bi-adjoint6}

In this section we will compute tree-level Witten diagrams for the bi-adjoint theory in AdS$_{6}$. For simplicity we will study the conformally-coupled scalar, i.e. we set $\nu=1/2$. The conformally-coupled, bi-adjoint theory in AdS$_{6}$ is particularly simple because at tree-level it is conformally invariant.

It is convenient to introduce a set of AdS Mandelstam-like invariants. First, from Lorentz invariance we know a general $n$-point correlation function will depend on $n(n-1)/2$ dot products of the momenta, $\bk_i\cdot \bk_j$, where $i,j=1,...,n-1$.\footnote{Conformal invariance imposes additional constraints on the functional form of the correlation function, but we will not use them.} We then define the AdS Mandelstam invariants to be:
\begin{align}
\label{eq: definition of s tilde}
\tilde s_{i_1\dots i_m}\equiv \left(
\sum\limits_{a=1}^m k_{i_a}+\abs{\sum\limits_{a=1}^m \bk_{i_a}}\right)
\left(
\sum\limits_{a=m+1}^n k_{i_a}+\abs{\sum\limits_{a=1}^m \bk_{i_a}}\right).
\nonumber \\
\end{align}
Our definition of the AdS Mandelstams is motivated by their connection to the usual Mandelstam invariants in the flat space limit. To see this, we first construct a null, $(d+1)$-dimensional momentum by appending the norm $k$ to the vector itself,
\begin{equation}
\label{eq: definition of k tilde}
\widehat{\bk}\equiv(ik,\bk).
\end{equation}
We then define the flat space invariants,
\begin{equation}
 s_{i_1\dots i_m}\equiv \left(\sum\limits_{a=1}^m \widehat{\bk}_{i_a}\right)\.\left(\sum\limits_{a=1}^m \widehat{\bk}_{i_a}\right)~.
 \end{equation}
At four-points we use the standard notation $s=s_{12}$, $t=s_{23}$, and $u=s_{13}$.
Finally, we recall that the flat space limit in AdS momentum space \cite{Raju:2012zr} is defined by analytically continuing in the norms $k_i$ such that the total ``energy'' $E_T\rightarrow 0$. Here $E_T$ is the sum of all the norms,
\be 
E_T\equiv \sum\limits_{i=1}^n k_i~.
\ee 
In this limit the AdS Mandelstam invariants go exactly to the corresponding flat space invariants if we identify the flat space null momenta as $\widehat \bk$:
\be 
\tilde s_{i_1\dots i_n}\;\;\xrightarrow{\text{flat space limit}} \;\;s_{i_1\ldots i_n}~.
\ee 

\begin{figure}
	\centering
	\includegraphics[scale=.8]{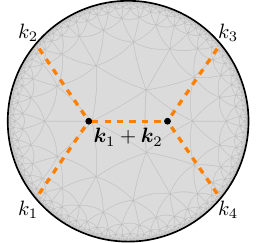}\qquad
	\includegraphics[scale=.8]{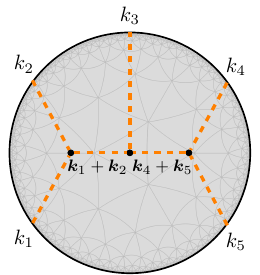}
	\caption{\label{fig: Four and five point exchange diagram for bi-adjoint scalar}Four and five point exchange diagram for bi-adjoint scalars.}
\end{figure}

Now we turn to computing Witten diagrams for the full, color-dressed correlator. To set the notation, we will use $M(1,2,3,4)$ to denote the color-dressed correlator and $A(1,2,3,4)$ for the color-ordered correlator. To keep the expressions compact, we will suppress the global symmetry indices. The exchange diagram for conformally coupled scalars in Figure~\ref{fig: Four and five point exchange diagram for bi-adjoint scalar} has been computed --- in \cite{Arkani-Hamed:2017fdk} for the dS wave function and in \cite{Albayrak:2020isk} for an AdS/CFT correlator --- so we will quote the final answer here:
\be 
\label{eq: AdS four point amplitude}
	M^{\text{sc}}(1,2,3,4)=\frac{(i\lambda)^2}{E_T}\left(\frac{n_sc_s}{\tilde{s}}+\frac{n_tc_t}{\tilde{t}}+\frac{n_uc_u}{\tilde{u}}\right)~,
\ee
where we use similar conventions as in flat space:
\be 
\tilde{s}=\tilde{s}_{12}\;,\quad
\tilde{t}=\tilde{s}_{23}\;,\quad
\tilde{u}=\tilde{s}_{13}~.
\ee
The s-channel color and ``kinematic'' factors are
\bea 
c_s=&f^{A_1A_2B}f^{A_3A_4B},
\\
n_s=&f^{a_1a_2b}f^{a_3a_4b},
\eea 
with repeated indices summed.
The $t$ and $u$-channel factors are defined by performing the following replacements:
\bea 
	n_tc_t=&n_sc_s\bigg|_{1\rightarrow 2\rightarrow 3\rightarrow 1}\;,\\ n_uc_u=&n_sc_s\bigg|_{1\rightarrow 3\rightarrow 2\rightarrow 1}\;.
\eea
As a consistency check, the AdS four-point function reduces to the familiar flat space amplitude when we take the residue at $E_T=0$:
\be 
M^{\text{sc}}(1,2,3,4)\xrightarrow{\text{flat space limit}} (i\lambda)^2\left(\frac{n_sc_s}{s}+\frac{n_tc_t}{t}+\frac{n_uc_u}{u}\right).
\ee

For the bi-adjoint theory, we have made an arbitrary split between the color and kinematic factors. Both $n_i$ and $c_i$ are by definition the group theory factors for $SU(N)$ and therefore obey the Jacobi identities:
\bea 
	n_s+n_t+n_u=&0\;,
	\\
	c_s+c_t+c_u=&0\;.
\eea 

While this example is trivial in terms of deriving a color-kinematics duality, it does demonstrate in a simple way how this duality will differ between AdS and flat space. To see this, we use the color relation $c_t=-c_s-c_u$ to write the color-dressed correlator as a sum of color-ordered correlators:
\begin{align}
	M^{\text{sc}}(1,2,3,4)=(i\lambda)^2 \bigg(&A^{\text{sc}}(1,2,3,4)c_s
	-A^{\text{sc}}(1,3,2,4)c_u\bigg)~,
\end{align}
where
\bea 
	A^{\text{sc}}(1,2,3,4)&=\frac{1}{E_T}\left(\frac{n_s}{\tilde{s}}-\frac{n_t}{\tilde{t}}\right),
	\\
	A^{\text{sc}}(1,3,2,4)&=\frac{1}{E_T}\left(\frac{n_t}{\tilde{t}}-\frac{n_u}{\tilde{u}}\right).
\eea 
We can further reduce this expression by using the identity $n_t=-n_s-n_u$ to find the following linear relations:
\bea[eq:solvKinRel] 
	A^{\text{sc}}(1,2,3,4)&=\frac{1}{E_T}\left(n_s\left(\frac{1}{\tilde{s}}+\frac{1}{\tilde{t}}\right)+\frac{n_u}{\tilde{t}}\right),
	\\
	A^{\text{sc}}(1,3,2,4)&=-\frac{1}{E_T}\left(n_u\left(\frac{1}{\tilde{u}}+\frac{1}{\tilde{t}}\right)+\frac{n_s}{\tilde{t}}\right).
\eea 
In flat space it is impossible to invert these equations and solve for the numerators directly in terms of the color-ordered amplitudes. This degeneracy implies a further identity among the color-ordered amplitudes, which are known as the BCJ relations. In flat space this degeneracy follows from the fact $s+t+u=0$ for massless scalars. However, in AdS we have $\tilde{s}+\tilde{t}+\tilde{u}\neq 0$ and we find:
\bea 
	n_s=&E_{T}\frac{\tilde{s}(\tilde{u}A^{\text{sc}}(1,3,2,4)+(\tilde{t}+\tilde{u})A^{\text{sc}}(1,2,3,4))}{\tilde{s}+\tilde{t}+\tilde{u}}~,
	\\
	n_u=&-E_{T}\frac{\tilde{u}(\tilde{s}A^{\text{sc}}(1,2,3,4)+(\tilde{s}+\tilde{t})A^{\text{sc}}(1,3,2,4))}{\tilde{s}+\tilde{t}+\tilde{u}}~.
\eea 
In the limit $E_T\rightarrow 0$ the explicit factor of $E_T$ cancels against the pole in the color-ordered correlator, while the sum in the denominator vanishes, $\tilde{s}+\tilde{t}+\tilde{u}\rightarrow 0$. Therefore, the kinematic numerators na\"ively diverge in the flat space limit.\footnote{When defining the numerators, we pulled out an overall power of $E_{T}^{-1}$, so while the full correlator diverges as $E_T\rightarrow0$, the numerators should remain finite.} To avoid this, we impose the following relation on the \textit{flat space} amplitude:
\be 
	sA^{\text{sc}}_{\text{flat}}(1,2,3,4)+(s+t)A^{\text{sc}}_{\text{flat}}(1,3,2,4)=0~,
\ee 
which one can recognize as the four-point BCJ relation. Alternatively, one can note that in the flat space limit \equref{eq:solvKinRel} becomes the standard flat space relations between the color-ordered amplitudes and the numerators.
If we take the same combination of color-ordered AdS correlators as what appears in the BCJ relation, we see the right hand  side is non-zero, but vanishes in the flat space limit:
\be 
	\tilde{s}A^{\text{sc}}(1,2,3,4)+(\tilde{s}+\tilde{t})A^{\text{sc}}(1,3,2,4)=-\frac{n_u}{E_T}\frac{\tilde{s}+\tilde{t}+\tilde{u}}{\tilde{u}}~.
\ee 

The observation that color-kinematics does not always yield BCJ relations has also been made in flat space.
For example, color-kinematic duality for massive, flat space amplitudes does not necessarily imply additional linear relations among the color-ordered amplitudes \cite{Momeni:2020vvr,Johnson:2020pny}. 
Instead, requiring that the BCJ relations hold --- or that there are only $(n-3)!$ linearly independent, color-ordered amplitudes at $n$-points --- imposes constraints on the masses of the particles. 
These constraints proved important in constructing valid examples of massive double-copy. 
Similar observations about color-kinematics and BCJ relations have also been made for ABJM \cite{Huang:2013kca,Sivaramakrishnan:2014bpa} and for amplitudes in the flat space bi-adjoint theory with off-shell momenta \cite{Du:2011js}. 

Finally, it is straightforward to generalize our results to higher points. For example, using the results of  \cite{Albayrak:2020isk}, we find that the five-point, color-dressed correlator is:
\begin{multline}
\label{eq: 5point expr}
\hspace{-.8em}M^{\text{sc}}(1,2,3,4,5)=-\frac{(i\lambda)^3}{E_T}\frac{n_s^{(5)}c_s^{(5)}}{\tilde s_{12}\tilde s_{123}}\left(1+\frac{E_T}{E_T+\w_{12}^-+\w_{45}^-}\right)\\+\text{crossed channels}~,
\end{multline}
where for brevity we defined
\be 
\label{eq: definition of omega}
\w^\pm_{i_1\dots i_m}\equiv 
\sum\limits_{a=1}^m k_{i_a}\pm \abs{\sum\limits_{a=1}^m \bk_{i_a}}\;.
\ee 
These objects are related to Mandelstam variables by 
\be 
\w^+_{i_1\dots i_m}=&\frac{\tilde s_{i_1\dots i_m}-s_{i_1\dots i_m}}{E_T}\;,\\
\w^-_{i_1\dots i_m}=&\frac{E_T s_{i_1\dots i_m} }{\tilde s_{i_1\dots i_m}-s_{i_1\dots i_m}}\;.
\ee 
In particular, $s_{i_1\dots i_m}=\w^+_{i_1\dots i_m}\w^-_{i_1\dots i_m}$. Finally, $n_s^{(5)}$ and $c_s^{(5)}$ are contractions of color structures which are defined explicitly in Appendix~\ref{sec: review of five point amplitudes}. 

One can also see that the AdS expression has the correct pole structure in the flat space limit:
\begin{multline}
	M^{\text{sc}}(1,2,3,4,5)\xrightarrow{\text{flat space limit}}-(i\lambda)^3\frac{n_s^{(5)}c_s^{(5)}}{ s_{12} s_{123}}\\+\text{crossed channels}\;.
\end{multline}

In Appendix~\ref{sec: review of five point amplitudes}, we give the expansion of the five-point, color-dressed correlator in terms of the color-ordered correlators. Color-kinematics in the bi-adjoint theory is trivial at all points and we find again that there is a square, non-degenerate matrix relating the color-ordered AdS correlators and the numerators. That is, if we organize the correlators and numerators into vectors, $A_\a$ and $n_\b$, we have the linear relation,
\begin{align}
A_\a=\sum\limits_{\b}S_{\a\b}n_{\b}~,
\end{align}
where the matrix $S$ is invertible.
Therefore, the 5-point numerators can be written as a linear combination of color-ordered correlators.
In the flat space limit, $\det S\rightarrow 0$ and one instead finds BCJ relations among the color-ordered, flat space amplitudes.

\subsection{Generalization to AdS$_{d+1}$}
\label{sec:bi-adjointd}
In this section we will study the bi-adjoint theory in general dimensions. The immediate difficulty one faces is that in generic dimensions, the exchange Witten diagrams for $\phi^3$ theory do not take a simple form, even for conformally coupled scalars. For example, in AdS$_4$ the exchange Witten diagrams already involve dilogarithms \cite{Arkani-Hamed:2015bza,Hillman:2019wgh,Albayrak:2020isk}. On the other hand, in flat space the tree-level amplitudes take a simple form in all dimensions, with poles corresponding to particle exchange. Therefore, for general AdS$_{d+1}$, it may not be clear how to identify the ``numerators'' which should obey the color-kinematic relations.

One remedy for this is to simply define a numerator $n_i$ as the overall coefficient of a tree-level exchange diagram whose color-factors have been removed. If we consider conformally-coupled, bi-adjoint scalars in AdS$_6$, this gives the same definition of the numerators as before, up to a factor of $E_T$.
Since we will not need the explicit form of the integrated diagram, in this section we will let the boundary scalars have a generic conformal dimension.

The color-dressed correlator for bi-adjoint scalars is
\be 
\label{eq: w's in bi adjoint scalar}
	M^{\text{sc}}(1,2,3,4)=n_sc_s W_s+n_tc_t W_t+n_uc_u W_u~,
\ee 
where the $s$-channel exchange Witten diagram, with the color-factors removed, is:
\begin{align}
W_{s}(\bk_i)=&(i\lambda)^2\int  \frac{dz_1dz_2}{(z_1z_2)^{d+1}}K_{\nu}^{\text{sc}}(k_1,z_1)K_{\nu}^{\text{sc}}(k_2,z_1)
\nonumber \\&G^{\text{sc}}_{\nu}\left(\abs{\bk_1+\bk_2},z_1,z_2\right)K_{\nu}^{\text{sc}}(k_3,z_2)K_{\nu}^{\text{sc}}(k_4,z_2)~.
\end{align}
The $t$ and $u$-channel diagrams are defined by the same permutations as before.

At this point we can repeat the analysis of the previous section with minor changes. Everywhere we see a factor of $(E_T\tilde{s})^{-1}$ we replace it with $W_{s}(\bk_i)$, and similarly for the $t$ and $u$-channel exchanges. Once again, we can use the color and kinematic relations to rewrite the $d$-dimensional color-ordered correlators in terms of the numerators, and then invert this relation. Suppressing the momentum arguments for compactness, we find:
\begin{align}
	n_s=\frac{{W}_{s}^{-1}}{W_s^{-1}+W_t^{-1}+W_u^{-1}} &\bigg[{W}_{u}^{-1}A^{\text{sc}}(1,3,2,4)
	\nonumber \\
	&\hspace{-1em}+(W_{t}^{-1}+W_u^{-1})A^{\text{sc}}(1,2,3,4)\bigg]~,
\end{align}
and
\begin{align}
	n_u=\frac{{W}_{u}^{-1}}{W_s^{-1}+W_t^{-1}+W_u^{-1}}& \bigg[{W}_{s}^{-1}A^{\text{sc}}(1,2,3,4)
	\nonumber \\
	&\hspace{-1em}+(W_{s}^{-1}+W_t^{-1})A^{\text{sc}}(1,3,2,4)\bigg]~.
\end{align}
Similarly, the BCJ relation in AdS becomes:
\begin{multline}
	W_s^{-1}A^{\text{sc}}(1,2,3,4)+(W_s^{-1}+W_{t}^{-1})A^{\text{sc}}(1,3,2,4)\\=-n_u\frac{W_s^{-1}+W_t^{-1}+W_u^{-1}}{W_u^{-1}}~.
\end{multline} 
In general dimensions, we do not have the explicit form of $W_{s,t,u}$ in AdS momentum space, although there do exist results in Mellin space \cite{Penedones:2010ue,Fitzpatrick:2011ia}. In our context, all we need is that in the flat space limit each Witten diagram becomes a flat space exchange diagram, e.g. \mbox{$W^{-1}_{s}\rightarrow s$}. We therefore see that the AdS BCJ relation has a non-zero right hand side which vanishes in the flat space limit.

Alternatively, we can define numerators by using the $p$ integral representation of the bulk-to-bulk propagator inside the Witten diagram,
\begin{align}
	W_s(\bk_i)=-i\int\frac{dz_1dz_2}{(z_1z_2)^{d+1}}\int\limits_{0}^{\infty}\frac{dp^2}{2}   K_{\nu}^{\text{sc}}(k_1,z_1) K_{\nu}^{\text{sc}}(k_2,z_1)
\nonumber \\ \frac{(z_1z_2)^{\frac{d}{2}}\cJ_{\nu}(pz_1)\cJ_{\nu}(pz_2)}{(\bk_1+\bk_2)^2+p^{2}-i\epsilon} K_{\nu}^{\text{sc}}(k_3,z_2)K_{\nu}^{\text{sc}}(k_4,z_2)~.
\end{align}

With this representation, we can use the flat space language and identify the numerators $n_i$ as multiplying certain poles in the momenta. The only difference is that in AdS we have a continuum of poles which depend on $p$. This ref\hspace{0.01cm}lects the well-known fact that in CFTs the generator of time translations has a continuous spectrum \cite{Rychkov:2016iqz}. For the bi-adjoint theory, the introduction of an integrand is not necessary since by definition the kinematic factors are independent of $p$. However, this representation will be useful once we turn to Yang-Mills in general dimensions.

\section{Yang-Mills Theory}
\label{sec:YM}
In this section we will study color-kinematics for Yang-Mills in AdS. The study of Yang-Mills in AdS$_{4}$ will mirror exactly the analysis of the bi-adjoint scalar theory in AdS$_{6}$. In both cases the theories are conformal at tree-level, and the correlators take similar forms. Then we will study Yang-Mills in general AdS$_{d+1}$ and propose how color-kinematics is manifested at the integrand level.

Throughout this section, we will take the axial gauge, $A_{z}=0$. With this choice, the propagators are \cite{Liu:1998ty,Raju:2011mp}:
\begin{multline}
G^{\text{YM}}_{\mu\nu}(\bk,z_1,z_2)=-i\int\limits_{0}^{\infty} \frac{dp^2}{2}(z_1z_2)^{\frac{d-2}{2}}\cJ_{\frac{d-2}{2}}(pz_1)
\\\frac{ \mathcal{T}_{\mu\nu}(\bk,p)}{\bk^{2}+p^{2}-i\epsilon}\cJ_{\frac{d-2}{2}}(pz_2)~,
\end{multline}
and
\be
K^{\text{YM}}(k,z)&=\frac{- i}{\Gamma(d/2)2^{d/2-1}}(kz)^{d/2-1}K_{\frac{d-2}{2}}(kz)~,
\ee
where the tensor $\mathcal{T}_{\mu\nu}(\bk,p)=\eta_{\mu\nu}+\frac{\bk_{\mu}\bk_{\nu}}{p^{2}}$.  Technically, the bulk-to-boundary propagator also comes with the tensor structure $\mathcal{T}_{\mu\nu}(\bk,\sqrt{-\bk^{2}})$, but this simply projects onto polarization vectors transverse to $\bk$. Throughout  this section we assume the polarizations are transverse, and therefore drop the projector.

\begin{figure}
	\centering
	$\begin{aligned}
		\begin{aligned}
			\includegraphics[scale=.65]{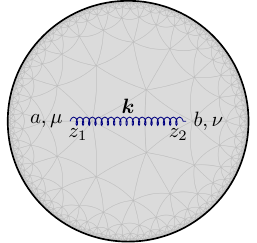}
		\end{aligned}=&\delta_{ab}G^{\text{YM}}_{\mu\nu}(\bk,z_1,z_2)\\
		\begin{aligned}
			\includegraphics[scale=.65]{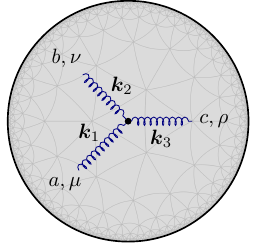}
		\end{aligned}=&z^4V^{abc}_{\mu\nu\rho}(\bk_1,\bk_2,\bk_3)\\
		\begin{aligned}
			\includegraphics[scale=.65]{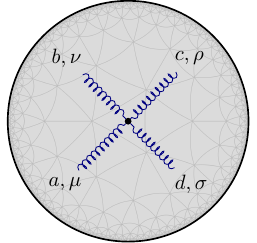}
		\end{aligned}=&z^4V^{abcd}_{\mu\nu\rho\sigma}
	\end{aligned}$
	\caption{\label{fig: Feynman rules for gauge field} Feynman rules for Yang-Mills theory in AdS.}
\end{figure}

The interaction terms in the axial gauge have the same momentum dependence as in flat space:
\be
	V^{abc}_{\mu\nu\rho}(\bk_i)=&gf^{abc}(\eta_{\mu\nu}(\bk_1-\bk_2)_\rho+\eta_{\nu\rho}(\bk_2-\bk_3)_{\mu}
\nn \\ &\hspace{.4in}+\eta_{\rho\mu}(\bk_3-\bk_1)_\nu)~,
\ee 
and 
\be 
	V^{abcd}_{\mu\nu\rho\sigma}=&-ig^{2}\bigg[c_s(\eta_{\mu\rho}\eta_{\nu\sigma}-\eta_{\mu\sigma}\eta_{\nu\rho})
	\nn\\
	&\hspace{.325in}+c_u(\eta_{\mu\sigma}\eta_{\rho\nu}-\eta_{\mu\nu}\eta_{\rho\sigma})
	\nn\\
	&\hspace{.325in}+c_t(\eta_{\mu\nu}\eta_{\sigma\rho}-\eta_{\mu\rho}\eta_{\sigma\nu})\bigg]~,
\ee
where the color factors are defined as before. To keep expressions compact, we will define:
\be 
	V^{abc}_{123}(\bk_1,\bk_2,\bk_3)=\epsilon_1^{\mu_1}\epsilon_2^{\mu_2}\epsilon_3^{\mu_3}V^{abc}_{\mu_1\mu_2\mu_3}(\bk_1,\bk_2,\bk_3)~,
\ee 
for the transverse polarization vectors $\e_i^{\mu_i}$, and similarly for the quartic interaction. We also raise and lower the $\mu,\nu$ indices using the flat space metric $\eta_{\mu\nu}$. To take into account that we are studying a theory in the Poincar\'e patch, we also need a factor of $z^4$ for each interaction vertex. We summarize the Feynman rules  in Figure~\ref{fig: Feynman rules for gauge field}.

\subsection{Test case: AdS$_{4}$}
\label{sec:YM4}
\begin{figure}
	\centering
	\includegraphics[scale=0.8]{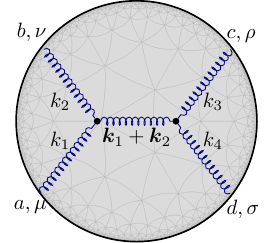}
	\caption{\label{fig: Four point exchange diagram for Yang-Mills theory}Four point exchange Witten diagram for Yang-Mills theory.}
\end{figure}

As we mentioned above, the advantage of studying Yang-Mills in AdS$_{4}$ is that the theory is conformal at tree-level. For example, the $s$-channel exchange diagram in Figure~\ref{fig: Four point exchange diagram for Yang-Mills theory} has been computed in both AdS \cite{Albayrak:2018tam, Albayrak:2019asr} and dS \cite{Baumann:2020dch} and takes the following simple form:\footnote{We note
that $\cW$ denotes the full Witten diagram of Yang-Mills, whereas $W$  in section~\ref{sec:bi-adjointd} denotes the Witten diagram of the bi-adjoint scalar with its color factor and numerator removed.}
\begin{align}
 \mathcal{W}^{\text{YM}}_{\text{s}}(\bk_i)= \frac{-ig^{2}c_s}{\tilde{s}E_T}&V^{12\mu}(\bk_1,\bk_2,-\bk_1-\bk_2) 
\nonumber \\
&\hspace{-1.5cm}\bigg(\eta_{\mu\nu}+\frac{(\abs{\bk_1+\bk_2}+E_T)(\bk_1+\bk_2)_\mu (\bk_1+\bk_2)_\nu}{(\abs{\bk_1+\bk_2})(k_1+k_2)(k_3+k_4)}\bigg)
 \nonumber \\
 &V^{34\nu}(\bk_3,\bk_4,\bk_1+\bk_2)~, \label{eq:YMsExch}
\end{align}
and the other exchange diagrams are found by permutation. The contact diagram is even simpler and is given by the flat space vertex times the total energy pole:
\be
	\mathcal{W}^{\text{YM}}_{\text{cont}}(\bk_i)=&\frac{V^{abcd}_{1234}}{E_T}~.\label{eq:YMcont}
\ee

Next, we want to rearrange the full color-dressed, Yang-Mills result into the form
\be 
	M^{\text{YM}}(1,2,3,4)=\frac{-ig^2}{E_T}\left(\frac{c_sn_s}{\tilde{s}}+\frac{c_tn_t}{\tilde{t}}+\frac{c_un_u}{\tilde{u}}\right)~.\label{eq:numYMDef}
\ee 
In order to do this, we follow the flat space prescription and split the contact diagram into three pieces, corresponding to the color structures $c_i$. For example, the $s$-channel piece of the contact diagram is
\be 
	\mathcal{W}^{\text{YM}}_{\text{cont},s}(\bk_i)=-\frac{i g^{2}}{E_T}c_s(\e_{13}\e_{24}-\epsilon_{14}\e_{23})~,
\ee 
where we defined 
\be 
\e_{ij}\equiv\e_i\.\e_j\;.
\ee 
Then to bring this term into the form \eqref{eq:numYMDef} we multiply $\mathcal{W}^{\text{YM}}_{\text{cont},s}$ by $\tilde{s}/\tilde{s}$, and similarly for the $t$ and $u$-channel pieces of the contact diagram. With these manipulations, we can bring the AdS correlator into the standard form \eqref{eq:numYMDef}, for which $n_s$ reads
\begin{align}
n_s=
&4\left(\bk_1\cdot \epsilon _2 \left(\epsilon _{13} \bk_3\cdot \epsilon _4-\epsilon
_{14} \bk_4\cdot \epsilon _3\right)- 1\leftrightarrow 2\right)
\nonumber \\
&-\left[4\epsilon _{12} \left(\bk_1\cdot \epsilon _3 \bk_3\cdot \epsilon _4-3\leftrightarrow 4\right) + (1\,\arr{.8em}\,4\,\arr{.8em}\, 2\,\arr{.8em}\, 3\,\arr{.8em}\, 1)\right]
\nonumber\\
&-\epsilon _{12} \epsilon _{34} (\widehat \bk_1-\widehat \bk_2)\cdot (\widehat \bk_4-\widehat \bk_3)+\tilde s (\epsilon _{13} \epsilon _{24}-\epsilon _{14} \epsilon _{23})
\nonumber\\
&+E_T\epsilon
_{12} \epsilon _{34}\frac{\left(k_1-k_2\right) \left(k_4-k_3\right)}{\abs{\bk_1+\bk_2}}~.
\end{align}
The term proportional to $\tilde{s}$ comes from the quartic interaction. As a reminder, the $\widehat{\bk}$ are the null flat space momenta, which we have used to make the expression more compact. In this form the AdS correlator does not obey the color-kinematic relations. One can check that \mbox{$n_s+n_t+n_u\neq0$} but that we have \mbox{$n_s+n_t+n_u\rightarrow 0$} in the flat space limit. The fact we have color-kinematics in this limit follows from the fact that the individual Witten diagrams have the correct flat space limit. Since color-kinematics holds automatically for flat space, four-point, Yang-Mills amplitudes, the corresponding AdS numerators must also obey color-kinematics in the limit $E_T\rightarrow 0$.

However, the AdS numerators are not unique and it is possible to define a new set of numerators related by a generalized gauge transformation:\footnote{The same transformation was found in \cite{Armstrong:2020woi}.}
\bea
	n'_{s}&= n_s-\tilde{s}\Omega~,
	\\
	n'_{t}&= n_t-\tilde{t}\Omega~,
	\\
	n'_{u}&= n_u-\tilde{u}\Omega~.
\eea
With these new numerators, the full correlator is unchanged:
\be 
	\frac{c_sn'_s}{\tilde{s}}+\frac{c_tn'_t}{\tilde{t}}+\frac{c_un'_u}{\tilde{u}}=\frac{c_sn_s}{\tilde{s}}+\frac{c_tn_t}{\tilde{t}}+\frac{c_un_u}{\tilde{u}}~,
\ee 
where we used the color Jacobi identity $c_s+c_t+c_u=0$. Therefore, if we choose:
\be 
	\Omega=\frac{n_s+n_t+n_u}{\tilde{s}+\tilde{t}+\tilde{u}}~,
\ee 
the new numerators $n'_s$ automatically satisfy color-kinematic duality. Here we see that our freedom in choosing a generalized gauge transformation such that the duality holds relies on having $\tilde{s}+\tilde{t}+\tilde{u}\neq0$. A similar observation at four-points was also seen for massive, flat space amplitudes \cite{Johnson:2020pny} and was made independently in \cite{Armstrong:2020woi}.

With this generalized gauge transformation, we can now repeat exactly the analysis we did for the conformal bi-adjoint scalar in AdS$_{6}$. By imposing both the color and kinematic identities, $c_s+c_t+c_u=0$ and $n'_s+n'_t+n'_u=0$, we can express the numerators in terms of the color-ordered, Yang-Mills correlators:\footnote{One can also check that the AdS color-ordered correlators satisfy the same $U(1)$ decoupling identity as their flat space counterparts.}
\begin{align}\label{eq: n's in terms of partial amplitudes}
	&n'_s=\frac{iE_T}{g^2}\frac{\tilde{s}}{\tilde{s}+\tilde{t}+\tilde{u}}
	\bigg(\tilde{u}A^{\text{YM}}(1,3,2,4)
	\nonumber\\ &\hspace{9em}
	+(\tilde{t}+\tilde{u})A^{\text{YM}}(1,2,3,4)\bigg),\\
	{}&n'_u=-\frac{iE_T}{g^2}\frac{\tilde{u}}{\tilde{s}+\tilde{t}+\tilde{u}}\bigg(\tilde{s}A^{\text{YM}}(1,2,3,4)
	\nonumber\\ &\hspace{9em}
	+(\tilde{s}+\tilde{t})A^{\text{YM}}(1,3,2,4)\bigg).
\end{align}
The only difference in comparison to the bi-adjoint scalar is that we had to shift the numerators in order for color-kinematics to hold. 

\subsection{Generalization to AdS$_{d+1}$}
\label{sec:YMd}
In this section we will study Yang-Mills in AdS$_{d+1}$. As with the bi-adjoint scalar, we face the problem that the gauge theory Witten diagrams are not known in closed form for general dimensions. For the bi-adjoint scalar, one solution was to simply express the full correlator as a sum of exchange diagrams. In the study of Yang-Mills in AdS, we face the additional challenge of the contact interaction, which we need to re-express such that it looks like a sum of exchange diagrams. Without a closed form expression in general dimensions, we can not simply multiply by $\tilde{s}/\tilde{s}$ to rewrite the contact diagram in this way.

Our resolution for this problem is to study the AdS Witten diagrams under the $p$ and $z$ integrals. We then want to understand color-kinematics at the level of the AdS integrand, which looks similar in structure to a flat space amplitude.

For example, using the explicit form of the cubic vertices, we find the exchange diagram is:
\begin{align}
\hspace{-.1cm}	\mathcal{W}^{\text{YM}}_{\text{exch,s}}(\bk_i)=-g^2c_s\int \hspace{-.15cm} \frac{dz_1dz_2}{(z_1z_2)^{d-3}}\mathfrak{t}_s^{\mu\nu} G^{\text{YM}}_{\mu\nu}(\bk_1+\bk_2,z_1,z_2)\Phi_{s},
\end{align}
where $\mathfrak{t}^{\mu\nu}$ is a product of three-point vertices,
\begin{align}
\mathfrak{t}_s^{\mu\nu}=&\left(\epsilon_{34}(\bk_4-\bk_3)^{\nu}-2\epsilon_3\cdot \bk_4\epsilon_4^\nu+2\epsilon_4\cdot \bk_3 \epsilon_3^{\nu}\right)
\nonumber
\\
&\left(\epsilon_{12}(\bk_1-\bk_2)^{\mu}+2\epsilon_1\cdot \bk_2\epsilon_2^\mu-2\epsilon_2\cdot \bk_1 \epsilon_1^{\mu}\right),
\end{align}
and $\Phi_{s}$ is the product of $s$-channel bulk-to-boundary propagators,
\begin{align}
\hspace{-0em}	\Phi_{s}=K^{\text{YM}}(k_1,z_1)K^{\text{YM}}(k_2,z_1)K^{\text{YM}}(k_3,z_2)K^{\text{YM}}(k_4,z_2).\nn \\
\end{align}
In analogy to flat space, these can be thought of as our external wavefunctions.

One difference in comparison to flat space is that our bulk-to-bulk propagator $G^{\text{YM}}_{\mu\nu}(k,z_1,z_2)$ is not proportional to the metric $\eta_{\mu\nu}$. Instead, we have extra factors which comes from our choice of axial gauge. The expression for the Witten diagram is simplest if we use the $p$ integral representation of the propagator: 
\begin{align}
\mathcal{W}^{\text{YM}}_{\text{exch,s}}(\bk_i)=ig^2c_s\int\hspace{-.1cm}  \frac{dz_1dz_2}{(z_1z_2)^{\frac{d-4}{2}}}\int\limits_{0}^{\infty} \frac{dp^{2}}{2}\mathfrak{t}_s^{\mu\nu}\mathcal{T}_{\mu\nu}(\bk_1+\bk_2,p)
\nonumber
\\
\frac{\mathcal{J}_{\frac{d-2}{2}}(pz_1)\mathcal{J}_{\frac{d-2}{2}}(pz_2)}{(\bk_1+\bk_2)^{2}+p^{2}-i\epsilon} \Phi_{s}~.
\end{align}
The way to interpret this expression, when making the analogy with flat space, is that the term
\begin{align}
	\widehat{G}_p^{\text{YM}}\left(\abs{\bk_1+\bk_2},z_1,z_2\right)=\frac{\mathcal{J}_{\frac{d-2}{2}}(pz_1)\mathcal{J}_{\frac{d-2}{2}}(pz_2)}{(\bk_1+\bk_2)^{2}+p^{2}-i\epsilon}~,
\end{align}
is the scalar piece of our propagator. Heuristically, we can think of $p$ as the radial momentum, although it only becomes a true component of the momentum in the flat space limit. 

For the $s$-channel piece of the contact diagram, we only have $z$-integrals:
\begin{multline}
	\cW^{\text{YM}}_{\text{cont,s}}(\bk_i)=-ig^2c_s\int \frac{dz}{z^{d-3}}(\epsilon_{13} \epsilon_{24}-\epsilon_{14}\epsilon_{23})\Phi_{s}~.
\end{multline}
To make this look like an exchange diagram, we introduce $p$-integrals via the following identity:
\begin{align}
	\int\limits_{0}^{\infty}\frac{dp^2}{2} J_{\nu}(pz_1)J_{\nu}(pz_2)=\frac{\delta(z_1-z_2)}{z_1}~.
\end{align}
The same identity was used in \cite{Raju:2011mp} to prove the validity of the BCFW recursion relations in AdS. We then find
\begin{align}
	\cW^{\text{YM}}_{\text{cont,s}}(\bk_i)=-ig^2c_s &\int \frac{dz_1dz_2}{(z_1z_2)^{\frac{d-4}{2}}}(\epsilon_{13} \epsilon_{24}-\epsilon_{14}\epsilon_{23})
	\nonumber
	\\ &\int\limits_{0}^{\infty} \frac{dp^2}{2}\mathcal{J}_{\frac{d-2}{2}}(pz_1)\mathcal{J}_{\frac{d-2}{2}}(pz_2)\Phi_{s}~.
\end{align}
It is now clear how to rewrite the contact diagram such that it looks like an exchange diagram, we multiply by $\frac{(\bk_1+\bk_2)^2+p^2}{(\bk_1+\bk_2)^2+p^2}$ under the integral. We now have the full $s$-channel piece of the color-dressed correlator:
\begin{align}
	M^{\text{YM}}_{s}=-ig^2 \int& \frac{dz_1dz_2}{(z_1z_2)^{\frac{d-4}{2}}}\int\limits_{0}^{\infty} \frac{dp^2}{2} c_sn_s 
	\nonumber\\
	&\widehat{G}_p^{\text{YM}}(\abs{\bk_1+\bk_2},z_1,z_2)\Phi_s~,
\end{align}
where the $s$-channel numerator is,
\begin{multline}
	n_s=(\epsilon_{13} \epsilon_{24}-\epsilon_{14}\epsilon_{23})((\bk_1+\bk_2)^{2}+p^{2})\\+\mathfrak{t}_s^{\mu\nu}\mathcal{T}_{\mu\nu}(\bk_1+\bk_2,p)~.
\end{multline}
Finally, the color-dressed correlator is the sum over the three-channels:
\be 
	M^{\text{YM}}(1,2,3,4)=M^{\text{YM}}_{s}+M^{\text{YM}}_{t}+M^{\text{YM}}_{u}~.
\ee

In order to bring all the numerators under one integral, one can switch to the position-space representation for the propagator and add plane-wave factors to the external wavefunctions $\Phi_{s,t,u}$. Then the full, color-dressed correlator is:
\be 
{}&M^{\text{YM}}(1,2,3,4)=-i g^2\int d^{d}x_1d^{d}x_2 \int \frac{dz_1dz_2}{(z_1z_2)^{\frac{d-4}{2}}}\\{}& \int\limits_{0}^{\infty} \frac{dp^2}{2}\widehat{G}_{p}^{\text{YM}}(x_1,z_1;x_2,z_2)\left(c_sn_s\widehat{\Phi}_s+c_tn_t\widehat{\Phi}_t+c_un_u\widehat{\Phi}_u\right)~,
\ee
where the new external wavefunctions are defined as:
\be 
	\widehat{\Phi}_s=e^{i( x_1\cdot(\bk_1+\bk_2)+x_2\cdot(\bk_3+\bk_4))}\Phi_s~.
\ee
While this representation nicely groups different terms together, we will find it convenient to work with the momentum space representation for the propagators. In general, one can also consider external wavefunctions which do not have translation invariance in the $d$ flat directions, in which case the $x$-integral representation is more useful.

Now we want to find a generalized gauge transformation such that the numerators obey color-kinematics duality, but the full correlator is left invariant. To do this, we define the following ``scalar'' exchange diagram for the $s$-channel,
\be 
	R_s(\bk_i,p)=-ig^2\int \frac{dz_1dz_2}{(z_1z_2)^{\frac{d-4}{2}}}\widehat{G}_{p}(\abs{\bk_1+\bk_2},z_1,z_2)\Phi_s~,
\ee
and similarly for the $t$ and $u$-channels.
If we define the shifted numerators as:
\bea 
	n'_s&=n_s-R_{s}^{-1}(\bk_i,p)\Omega(\bk_i,p)~,
	\\
	n'_t&=n_t-R_{t}^{-1}(\bk_i,p)\Omega(\bk_i,p)~,
	\\
	n'_u&=n_u-R_{u}^{-1}(\bk_i,p)\Omega(\bk_i,p)~,
\eea 
then $M_s$ shifts as:
\be 
	M^{\text{YM}}_s\rightarrow M^{\text{YM}}_s+c_s\int\limits_{0}^{\infty} \frac{dp^2}{2} \ \Omega(\bk_i,p)~.
\ee 
Therefore, as long as $\int_{0}^{\infty} \frac{dp^2}{2}\Omega(\bk_i,p)$ is finite, the color Jacobi identity guarantees this redefinition leaves the full correlator invariant. We find that the shifted numerators $n'_i$ satisfy color-kinematics duality if we set:
\be 
	\Omega(\bk_i,p)=\frac{n_s+n_t+n_u}{R_{s}^{-1}(\bk_i,p)+R_{t}^{-1}(\bk_i,p)+R_{u}^{-1}(\bk_i,p)}~.
\ee 
As an example, in AdS$_{4}$ we have:
\begin{multline}
	R_s(\bk_i,p)=-\frac{2 i p}{\pi  \left((\bk_1+\bk_2)^2+p^2\right)} \\ \frac{1}{ \left((k_1+k_2)^2+p^2\right) \left((k_3+k_4)^2+p^2\right)}~,
\end{multline}
and one can check that $\int_{0}^{\infty} \frac{dp^2}{2}\Omega(\bk_i,p)$ is finite.

We should emphasize, unlike in our previous analysis for Yang-Mills in AdS$_4$ or the bi-adjoint scalar in AdS$_{d+1}$, here the color-kinematic numerators are functions of both $p$ and $k$. Therefore, we cannot directly express the numerators in terms of the integrated, color-ordered correlators. It would be interesting if  there is another formulation of AdS color-kinematics where such relations hold in general $d$. It would also naturally be interesting to find a representation, e.g., in Mellin or position space, where the numerators are not directly expressible in terms of the AdS color-ordered correlators, and instead there are new BCJ-like relations for AdS correlators.

\section{Color KLT and Double Copy}
\label{sec:KLT}
In this section we will study simple examples of the KLT and double-copy relations in AdS. The simplest double-copy and KLT relations roughly state that \mbox{YM=YM $\otimes$ bi-adjoint}. More precisely, the color-dressed Yang-Mills correlator can be expressed as a product of color-ordered Yang-Mills and bi-adjoint correlators \cite{Bern:1999bx,Du:2011js}.\footnote{For an example of KLT in cosmology, see \cite{Li:2018wkt}.} 
We then discuss double-copy for gravity at 3- and 4-points.

To start, we can consider Yang-Mills in AdS$_4$, where the color-dressed correlator has the form:
\be 
	M^{\text{YM}}(1,2,3,4)=\frac{-ig^2}{E_T}\left(\frac{c_sn_s}{\tilde{s}}+\frac{c_tn_t}{\tilde{t}}+\frac{c_un_u}{\tilde{u}}\right)~.\label{eq:YM_CK}
\ee
This is a simple example of double-copy because we can think of $c_i$ as the numerators for the bi-adjoint theory and $n_i$ are of course the numerators of the Yang-Mills theory. Assuming color-kinematics holds for the Yang-Mills numerators, we can directly express them in terms of the color-ordered Yang-Mills correlators. Similarly, we can also express the color-factors $c_i$ in terms of the color-ordered, bi-adjoint correlators. To find a KLT relation, we then write each of the numerators in terms of the corresponding color-ordered correlators.

There are two important differences in comparison to the flat space, color KLT relations. The first is that here the AdS KLT matrix has rank two. This is expected because in AdS we have two linearly independent color-ordered correlators at four-points, while in flat space we only have one independent amplitude. The second is that we have an extra degree of freedom: when writing the color factors $c_i$ in terms of the bi-adjoint correlators, we are free to choose the spacetime dimension $d$. In flat space we have a similar freedom, but there the scalar amplitudes look the same in all dimensions, while in AdS the form can change dramatically. For example, while we take the Yang-Mills theory to live in AdS$_{4}$, we are free to express $c_i$ in terms of the conformally-coupled, bi-adjoint scalar theory in AdS$_{6}$. With this choice the color KLT relation takes the form:
\begin{multline}
M^{\text{YM}}(1,2,3,4)=\\ \left(A^{(d)}_{\text{sc}}(1,2,3,4),A^{(d)}_{\text{sc}}(1,3,2,4)\right)^{T}\cdot K^{(d,3)} \\\cdot  \left(A^{(d=3)}_{\text{YM}}(1,2,3,4),A^{(d=3)}_{\text{YM}}(1,3,2,4)\right).
\end{multline}
Here the superscripts in $K^{(d_1,d_2)}$ give us the dimension of the AdS$_{d_i+1}$ spacetime in which the bi-adjoint scalar and the Yang-Mills theory live, respectively.\footnote{ For $d_1=5$ and $d_2=3$, we have
	\be 
	{}&K^{(5,3)}=\frac{E_T^2}{\tilde{t}(\tilde{s}+\tilde{t}+\tilde{u})^{2}}\\ {}&
	\begin{pmatrix}
		\tilde{s} (\tilde{t}+\tilde{u}) (2 \tilde{s} \tilde{u}+\tilde{t} (\tilde{t}+\tilde{u})) & \tilde{s} \tilde{u} (2 \tilde{s}+3 \tilde{t}) (\tilde{t}+\tilde{u}) \\
		\tilde{s} \tilde{u} (\tilde{s} (\tilde{t}+2 \tilde{u})+2 \tilde{t} (\tilde{t}+\tilde{u})) & \tilde{u} \left(\tilde{u} (2 \tilde{s}+\tilde{t}) (\tilde{s}+\tilde{t})+\tilde{s} \tilde{t} \tilde{u}+\tilde{t} (\tilde{s}+\tilde{t})^2\right)
	\end{pmatrix}\;.
	\ee 
} The AdS KLT matrix becomes singular in the flat space limit, reflecting the additional linear relations for flat space amplitudes. Here we restricted Yang-Mills to AdS$_4$, so that its integrated correlator took a simple form, but it would be interesting to extend this discussion to integrated correlators in general dimensions.

Next, we will study how double-copy may work at the integrand level for general dimensions and for gravity. 
Below, we reproduce the $s$-channel piece of the AdS Yang-Mills integrand:
\begin{multline}
	M^{\text{YM}}_{s}=-i \int\frac{dz_1dz_2}{(z_1z_2)^{\frac{d-4}{2}}}\int\limits_{0}^{\infty} \frac{dp^2}{2} c_sn_s \\\widehat{G}_p^{\text{YM}}(\abs{\bk_1+\bk_2},z_1,z_2)\Phi_s~.
\end{multline} 
Given this expression, we can double-copy down to the bi-adjoint scalar by taking $n_s$ and replacing it with a $SU(N)$ color factor $c'_s$. This yields an exchange Witten diagram for a scalar in AdS$_{d+3}$ dual to a boundary scalar of dimension $\Delta=d$:
\begin{multline}
	M^{\text{sc}}_{\Delta=d,s}\bigg|_{d'=d+2}=-i \int \frac{dz_1dz_2}{(z_1z_2)^{\frac{d-4}{2}}}\int\limits_{0}^{\infty} \frac{dp^2}{2} c_sc'_s\\\widehat{G}_p^{\text{YM}}(\abs{\bk_1+\bk_2},z_1,z_2)\Phi_s~.
\end{multline}
The shift \mbox{$d\rightarrow d+2$} and the identification \mbox{$\Delta=d$} follows from matching this expression with the integrand for a scalar exchange diagram in AdS. Specifically, to find the dimension of the AdS spacetime and the conformal dimension of the scalar, we match the arguments of the Bessel functions and the overall powers of $z$.
As a consistency check, if we set $d=3$ we find a scalar of dimension $\Delta=3$ in AdS$_{6}$, i.e. the conformally-coupled scalar. By comparing \equref{eq: AdS four point amplitude} and \equref{eq:YM_CK}, we see explicitly that making the replacement $n_s\rightarrow c'_s$ for the AdS$_4$ Yang-Mills correlator gives the AdS$_6$ bi-adjoint scalar correlator, up to overall factors such as the couplings.

It is tempting to conjecture that if we replace $c_s$ in the Yang-Mills integrand with $n_s$ we get the $s$-channel contribution to graviton four-point scattering in AdS$_{d-1}$:\footnote{Recall that to compare the graviton amplitude with the double-copied Yang-Mills amplitude, we also need to write its contact term so that it looks like the sum of three exchanges.}
\begin{multline}
	M^{\text{GR}}_{s}\bigg|_{d'=d-2}\stackrel{?}{=}-i \int \frac{dz_1dz_2}{(z_1z_2)^{\frac{d-4}{2}}}\int\limits_{0}^{\infty} \frac{dp^2}{2} n_s^2\\\widehat{G}_p^{\text{YM}}(\abs{\bk_1+\bk_2},z_1,z_2)\Phi_s~.\label{eq:4ptgraviton}
\end{multline}
The shift from \mbox{$d\rightarrow d-2$} is once again found by comparing the $z$ dependence of the resulting expression to that of the graviton propagators, which we will give explicitly in a moment.\footnote{In comparison to \cite{Armstrong:2020woi}, here we perform the squaring under the $p$ and $z$ integral, i.e. directly for the integrand.} 
Here we assume that the AdS theory is given by Einstein gravity.

Unfortunately \eqref{eq:4ptgraviton} is inconsistent with the AdS cutting rules \cite{Meltzer:2020qbr}.\footnote{We thank Daniel Baumann for bringing this to our attention.}
The fact double-copy in AdS will have to be more complicated than a simple squaring may not be surprising. 
We have already seen that at tree-level the AdS gauge-boson exchange diagram is more complex than the corresponding flat space one and there is a similar increase in complexity for graviton diagrams \cite{Raju:2012zs,Albayrak:2019yve,Baumann:2020dch}. 
Furthermore, in flat-space squaring color-kinematic numerators is motivated by the fact the resulting amplitude obeys the graviton Ward identities \cite{Bern:2019prr}.
Ward identities in AdS/CFT are more complicated, owing to contact terms in the CFT correlators, and it would be interesting to understand what relations have to be imposed on gauge-theory numerators such that the double-copied correlator obeys the graviton, or stress-tensor, Ward identities. 

With these caveats in mind, one motivation to study AdS double-copy comes from relations between gauge and graviton scattering at three-points. The three-point correlator for Yang-Mills in AdS$_{d+1}$ is
\begin{multline}
	M_3^{\text{YM}}=g f^{abc}\int\frac{dz}{z^{d+1}}z^4\left(\epsilon_1\cdot (\bk_2-\bk_3)\epsilon_{2}\.\e_3+\text{cyclic}\right)\\\prod\limits_{i=1}^{3}K^{\text{YM}}(k_i,z)~,
\end{multline}
while the three-point correlator for Einstein gravity in AdS$_{d+1}$ is \cite{Raju:2011mp,Raju:2012zr,Albayrak:2019yve}
\begin{multline}
	M_3^{\text{GR}}=\sqrt{G_{N}}\int\frac{dz}{z^{d+1}}z^8\bigg(-2\epsilon_1\cdot \bk_2\epsilon_2\cdot \bk_3\epsilon_{1}\.\e_3\epsilon_{2}\.\e_3\\+\epsilon_1\cdot \bk_2\epsilon_1\cdot \bk_3(\epsilon_{2}\.\e_3)^2+\text{perms}\bigg)\prod\limits_{i=1}^{3}K^{\text{GR}}(k_i,z)~.
\end{multline}
Here we wrote the graviton polarization tensor as a product of null polarization vectors, $\epsilon_{\mu\nu}=\epsilon_{\mu}\epsilon_{\nu}$.
We also need the graviton bulk-to-boundary propagator:
\be 
	K^{\text{GR}}(k,z)=\frac{-i}{2^{d/2}\Gamma(d/2+1)}k^{d/2}z^{d/2-2}\cK_{d/2}(kz)~.
\ee 
As with the Yang-Mills bulk-to-boundary propagator, we have dropped an overall tensor structure which projects out polarizations along the momenta $\bk$.

Then, if we define the three-point numerator to be
\be
	n_{\text{3-pt}}=\epsilon_1\cdot (\bk_2-\bk_3)\epsilon_{2}\cdot \epsilon_3+\text{cyclic}~,
\ee 
we find
\be 
	M^{\text{GR}}_3\bigg|_{d'=d-2}\propto\int\frac{dz}{z^{d+1}}z^4n_{\text{3-pt}}^2\prod\limits_{i=1}^{3}K^{\text{YM}}(k_i,z)~.
\ee 
In other words, squaring the three-point numerator for Yang-Mills in AdS$_{d+1}$ yields the three-point correlator for Einstein gravity in AdS$_{d-1}$, up to some overall convention dependent factors. Alternatively, one can square the numerator and modify by hand the $z$-dependence so that the double-copied correlator also comes from gravity in AdS$_{d+1}$. 

\section{Conclusion}
\label{sec:Conclusion}

In this work we explored the viability of color-kinematics and double-copy in AdS momentum space. We found that color-kinematics for AdS four-point functions appears trivial, one can always perform a generalized gauge transformation such that the duality is valid. We also found that it is possible to express the numerators directly in terms of the color-ordered correlators and that the BCJ relations are modified by an extra term which vanishes in the flat space limit. We used the relation between the numerators and integrated correlators to find the AdS color KLT relation and discuss how double-copy in AdS may work at the integrand level.

There is clearly more work that needs to be done on this subject. In this note we focused on AdS momentum space because it has a natural connection to the wave function of the universe in cosmological spacetimes. There has also been recent beautiful work on the relation between momentum space correlators in AdS and dS and a new set of cosmological polytopes \cite{Arkani-Hamed:2017fdk,Arkani-Hamed:2018bjr,Benincasa:2018ssx,Benincasa:2019vqr,Benincasa:2020aoj}. 
For color-kinematics however, it could turn out that another representation is more useful, including twistor formulations \cite{Adamo:2012nn,Skinner:2013xp,Adamo:2013tja,Adamo:2015ina,Adamo:2016rtr}, spinor-helicity in stereographic coordinates \cite{Nagaraj:2018nxq,Nagaraj:2019zmk,Nagaraj:2020sji}, Mellin space \cite{Penedones:2010ue}, or of course position space.
Recent work on the scattering equation formalism \cite{Cachazo:2013gna,Cachazo:2013hca,Cachazo:2013iea} generalized to AdS \cite{Eberhardt:2020ewh,Roehrig:2020kck} will also prove invaluable in studying color-kinematics and double-copy in AdS. Based on related results for massive scattering amplitudes \cite{Johnson:2020pny}, we expect it is important to find a representation of AdS/CFT correlators such that color-kinematics, plus some possible assumptions on the spectrum, implies additional relations for the color-ordered correlators.

In flat space, color-kinematic duality and the double-copy relations extend to theories other than gauge or gravity theories. 
For instance, the nonlinear sigma model has been studied in \cite{Chen:2013fya,Carrasco:2016ygv}. 
Also, it was shown that the Lagrangian of the nonlinear sigma model exhibits a manifest duality between color and kinematics \cite{Cheung:2016prv}. 
It would be interesting to study these theories in AdS and see if color-kinematics can be understood at the Lagrangian level. 
Finally, there has been progress in computing loop level AdS correlators through bulk and boundary unitarity methods 
\cite{Fitzpatrick:2011dm,Aharony:2016dwx,Caron-Huot:2017vep,Ponomarev:2019ofr,Meltzer:2019nbs,Meltzer:2020qbr,Costantino:2020vdu}.\footnote{See also \cite{Goodhew:2020hob,Cespedes:2020xqq} for related developments in dS.} 
In flat space, generalized unitarity and double-copy relations can be systematically used to study higher-loop graviton amplitudes and reveal new ultraviolet cancellations \cite{Bern:2018jmv}. Loop computations in AdS is in its infancy in comparison to its flat space counterpart and it is conceivable that color-kinematics and double-copy could present a new way to study AdS loops.
\begin{acknowledgments}
We thank Daniel Baumann, Julio Parra-Martinez and Allic Sivaramakrishnan for discussions. We also thank Allic Sivaramakrishnan for comments on the draft. SA is supported by DOE grant no. DE-SC0020318 and Simons Foundation grant 488651 (Simons Collaboration on the Nonperturbative Bootstrap). The research of DM is supported by Simons Foundation grant 488657, the Walter Burke Institute for Theoretical Physics and the Sherman Fairchild Foundation.
\end{acknowledgments}

\appendix 
\renewcommand{\thesection}{\Alph{section}}
\numberwithin{equation}{section}
\section{Bi-adjoint, five point correlators}
\label{sec: review of five point amplitudes}
Here we will study the five-point color-dressed correlator for the conformally coupled, bi-adjoint scalar in AdS$_6$.\footnote{In this section we will drop the ``sc'' superscript as we will only study the five-point function for the bi-adjoint scalar.}
The color-dressed correlator can be written as a sum over $15$ exchange diagrams:
\be 
\label{eq: full 5point amplitude}
M(1,2,3,4,5)=\mathit{c}_{12345}n_{12345}W_{12345}+\text{crossed-channels}~,
\ee
where the color factors are defined as
\be 
c_{ijklm}\equiv f^{ij\a}f^{\a k\b}f^{\b lm}~.
\ee 
The $n_{ijklm}$ are defined in the same way, but for the second $SU(N)$ global symmetry. $W_{ijklm}$ is the Witten diagram in the corresponding channel with the color and kinematic factors removed. We follow the same ordering as in figure \ref{fig: Four and five point exchange diagram for bi-adjoint scalar}. The explicit expression for the five-point Witten diagram is
\be 
W_{12345}=-\frac{(i\lambda)^3}{E_T}\frac{1}{\tilde s_{12}\tilde s_{123}}\left(1+\frac{E_T}{E_T+\w_{12}^-+\w_{45}^-}\right)~,
\ee
where $\omega^{\pm}$ are defined in \equref{eq: definition of omega}. 

For completeness, the 15 diagrams are given by:
\begin{equation}
\begin{aligned}
	\{&W_{12345},W_{12435},W_{12534},W_{13245},W_{23145},
	\nonumber\\ & W_{32415},W_{32514},W_{42315},W_{42531},W_{43125},
	\nonumber\\ &W_{43215},W_{52341},W_{52431},W_{53142},W_{53241}\}~.
\end{aligned}
\end{equation}

There are nine independent Jacobi identities among the color factors:
\be
0=&\mathit{c}_{12435}-\mathit
{c}_{53142}+\mathit{c}_{53241}~,\\
0=&\mathit{c}_{13245}-\mathit{c}_{42531}+\mathit{c}_{52431}~,
\\0=&\mathit{c}_{43125}-\mathit{c}_{52341}
+\mathit{c}_{52431}~,
\\0=&\mathit{c}_{42315}+\mathit{c}_{42531}-\mathit{c}_{53142}~,
\\0=&\mathit{c}_{32514}+\mathit{c}_{52341}-\mathit{c}_{53241}~,
\\0=&\mathit{c
}_{32415}+\mathit{c}_{42531}+\mathit{c}_{43215}-\mathit{c}_{53142}~,
\\
0=&\mathit{c}_{12534}-\mathit{c}_{43215}+\mathit{c}_{52341}-\mathit{c}_{52431}~,
\\
0=&\mathit{c}_{23145}-\mathit{c}_{42531}-\mathit{c}_{43
	215}+\mathit{c}_{52341}+\mathit{c}_{53142}-\mathit{c}_{53241}~,
\\0=&\mathit{c}_{12345}+\mathit{c}_{43215}-\mathit{c}_{52341}+
\mathit{c}_{52431}-\mathit{c}_{53142}+\mathit{c}_{53241}~.
\ee
By using these identities, we can rewrite \equref{eq: full 5point amplitude} in terms of $6$ color-ordered correlators:
\be 
M(1,2,3,4,5)&= \mathit{c}_{12435}A(1,2,4,3,5)+\mathit{c}_{12534}A(1,2,5,3,4)\\&+ \mathit{c}_{32514}A(3,2,5,1,4)+\mathit{c}_{42531}A(4,2,5,3,1)
\\&+ \mathit{c}_{43125}A(4,3,1,2,5)+\mathit{c}_{52431}A(5,2,4,3,1)
\label{eq:5ptColorOrdDecomp}
~,
\ee 
where we made an arbitrary choice for the independent set of color factors. The color-ordered correlators can then be expressed in terms of the numerators $n_{ijklm}$. For example, we have:
\begin{align}
A(1,2,4,3,5) \ = \ &n_{12345}W_{12345}+n_{12435}W_{12435}
\nonumber \\-\ &n_{23145}W_{23145}+n_{32415}W_{32415}
\nonumber \\+\ &n_{42315}W_{42315}+n_{53142}W_{53142}~.
\end{align}
Similar relations can be found for the other 5 color-ordered correlators by comparing \equref{eq: full 5point amplitude} and \equref{eq:5ptColorOrdDecomp}, or equivalently by using the color-ordered Feynman rules \cite{Bern:2019prr}.
Using that the numerators $n_{ijklm}$ also obey the Jacobi relations, we can relate the 6 independent, color-ordered correlators to 6 independent numerators. If we organize the numerators and color-ordered correlators into vectors,
\be 
A\equiv 
(
	&A(1,2,4,3,5), \ \
	A(1,2,5,3,4), \ \
	A(3,2,5,1,4),\ \
	\\
	&A(4,2,5,3,1),\ \
	A(4,3,1,2,5),\ \
	A(5,2,4,3,1)
)
\ee
and
\be
\mathit{n}\equiv 
\begin{pmatrix}
	\mathit{n}_{12435} &
	\mathit{n}_{12534} &
	\mathit{n}_{32514} &
	\mathit{n}_{42531} &
	\mathit{n}_{43125} &
	\mathit{n}_{52431}
\end{pmatrix}
\ee 
Then the color-dressed correlator can be written as
\be 
M(1,2,3,4,5)=A_\a \mathit{c}_\a=S_{\a\b}\mathit{c}_\a\mathit{n}_\b
\ee 
for some matrix $S_{\a\b}$. In flat space, the corresponding matrix is degenerate due to the flat space BCJ relations. In AdS, $S_{\a\b}$ instead is a full-rank matrix, which can be checked using the explicit form of the five-point Witten diagram. We then find,
\be 
\mathit{n}=S^{-1}\.A~.
\ee 
This is the generalization of \equref{eq: n's in terms of partial amplitudes} to $5-$point amplitudes.

The explicit expression for $S_{\a\b}$ reads as 
\scriptsize
\be 
\label{eq: similarity matrix}
S_{1\b}=&\begin{pmatrix}
	W_{12345}+W_{12435}+W_{23145}+W_{32415}+W_{42315}+W_{53142} \\
	-W_{12345}-W_{23145}-W_{32415} \\
	W_{32415}+W_{42315}+W_{53142} \\
	-W_{23145}-W_{32415}-W_{42315} \\
	W_{42315}+W_{53142} \\
	W_{23145}+W_{32415}+W_{42315}+W_{53142} 
\end{pmatrix}
\\
S_{2\b}=&\begin{pmatrix}
	-W_{12345}-W_{23145}-W_{32415} \\
	W_{12345}+W_{12534}+W_{23145}+W_{32415}+W_{43215} \\
	-W_{32415} \\
	W_{23145}+W_{32415} \\
	W_{43215} \\
	-W_{23145}-W_{32415}
\end{pmatrix}
\\
S_{3\b}=&\begin{pmatrix}
	W_{32415}+W_{42315}+W_{53142} \\
	-W_{32415} \\
	W_{32415}+W_{32514}+W_{42315}+W_{53142}+W_{53241} \\
	-W_{32415}-W_{42315} \\
	W_{42315}+W_{53142}+W_{53241} \\
	W_{32415}+W_{42315}+W_{53142}+W_{53241} 
\end{pmatrix}
\\
S_{4\b}=&\begin{pmatrix}
	-W_{23145}-W_{32415}-W_{42315} \\
	W_{23145}+W_{32415} \\
	-W_{32415}-W_{42315} \\
	W_{13245}+W_{23145}+W_{32415}+W_{42315}+W_{42531} \\
	-W_{42315} \\
	-W_{13245}-W_{23145}-W_{32415}-W_{42315} 
\end{pmatrix}
\\
S_{5\b}=&\begin{pmatrix}
	W_{42315}+W_{53142} \\
	W_{43215} \\
	W_{42315}+W_{53142}+W_{53241} \\
	-W_{42315} \\
	W_{42315}+W_{43125}+W_{43215}+W_{52341}+W_{53142}+W_{53241} \\
	W_{42315}+W_{52341}+W_{53142}+W_{53241}
\end{pmatrix}
\\
S_{6\b}=&\begin{pmatrix}
	W_{23145}+W_{32415}+W_{42315}+W_{53142} \\
	-W_{23145}-W_{32415} \\
	W_{32415}+W_{42315}+W_{53142}+W_{53241} \\
	-W_{13245}-W_{23145}-W_{32415}-W_{42315} \\
	W_{42315}+W_{52341}+W_{53142}+W_{53241} \\
\left(\begin{aligned}
	W_{13245}+W_{23145}+W_{32415}+W_{42315}+W_{52341}\\+W_{52431}+W_{53142}+W_{53241}
\end{aligned}\right)
\end{pmatrix}
\ee 
\normalsize

\bibliography{CK_Draft}
\bibliographystyle{utphys}
\end{document}